\documentclass[showkeys,superscriptaddress,amsmath,amssymb,aps,prl,twocolumn]{revtex4-1}

\usepackage{graphicx}
\usepackage{dcolumn}
\usepackage{bm}
\usepackage{xcolor}
\usepackage{amsmath}
\usepackage{chngcntr}
\usepackage{nicefrac}
\usepackage{hyperref}
\usepackage{natbib}

\begin{document}
{

\title{Chemical and optical control of chiral-domain dynamics in 1\textit{T}-TaS$_2$}

\author{Qingzheng Qiu}\thanks{These authors contributed equally to this work.}
\affiliation{International Center for Quantum Materials, School of Physics, Peking University, Beijing 100871, China}

\author{Mengxian Zhao}\thanks{These authors contributed equally to this work.}
\affiliation{Beijing National Laboratory for Condensed Matter Physics and Institute of Physics, Chinese Academy of Sciences, Beijing 100190, China}
\affiliation{University of Chinese Academy of Sciences, Beijing 100049, China}

\author{Roman Mankowsky}
\author{Henrik Till Lemke}
\author{Serhane Zerdane}
\author{Mathias Sander}
\affiliation{Swiss Light Source, Paul Scherrer Institute, Villigen, Switzerland}

\author{Zihao Tao}
\author{Qizhi Li}
\affiliation{International Center for Quantum Materials, School of Physics, Peking University, Beijing 100871, China}

\author{Xiquan Zheng}
\affiliation{International Center for Quantum Materials, School of Physics, Peking University, Beijing 100871, China}

\author{Shilong Zhang}
\affiliation{International Center for Quantum Materials, School of Physics, Peking University, Beijing 100871, China}

\author{Qian Xiao}
\affiliation{International Center for Quantum Materials, School of Physics, Peking University, Beijing 100871, China}

\author{Xinyi Jiang}
\affiliation{International Center for Quantum Materials, School of Physics, Peking University, Beijing 100871, China}

\author{Xin Liu}
\author{Shih-Wen Huang}
\affiliation{Swiss Light Source, Paul Scherrer Institute, Villigen, Switzerland}

\author{Yang Yang}
\affiliation{Beijing National Laboratory for Condensed Matter Physics and Institute of Physics, Chinese Academy of Sciences, Beijing 100190, China}
\affiliation{University of Chinese Academy of Sciences, Beijing 100049, China}

\author{Sheng Meng}
\email{smeng@iphy.ac.cn}
\affiliation{Beijing National Laboratory for Condensed Matter Physics and Institute of Physics, Chinese Academy of Sciences, Beijing 100190, China}
\affiliation{University of Chinese Academy of Sciences, Beijing 100049, China}
\affiliation{Songshan Lake Materials Laboratory, Dongguan, Guangdong 523808, China}

\author{Yingying Peng}
\email{yingying.peng@pku.edu.cn}
\affiliation{International Center for Quantum Materials, School of Physics, Peking University, Beijing 100871, China}
\affiliation{Collaborative Innovation Center of Quantum Matter, Beijing 100871, China}

\date{\today}

\begin{abstract}

Optical control of symmetry-breaking quantum phases is often constrained when one domain is strongly favored in equilibrium. This limitation is exemplified by the chiral charge-density-wave (CDW) order in 1\textit{T}-TaS$_2$, where pristine samples predominantly select a single chirality. Here we show that free-energy landscape engineering through Ti substitution enables a distinct nonthermal pathway for ultrafast chiral-domain redistribution. Ti doping stabilizes coexisting chiral domains and tunes their relative stability, allowing femtosecond excitation to drive an asymmetric and anisotropic redistribution from the dominant toward the minority chirality. The subpicosecond response follows a $\sim2$~THz amplitude mode and is consistent with a phonon-assisted pathway involving transient domain-wall configurations. Our results establish free-energy landscape engineering as a strategy for selecting nonequilibrium transition pathways.

\end{abstract}

\maketitle

Chiral order in quantum materials can generate unconventional transport and nonreciprocal electronic responses~\cite{tokura2018nonreciprocal,nagaosa2024nonreciprocal}, motivating efforts to control chiral symmetry breaking on ultrafast timescales. The layered transition-metal dichalcogenide 1\textit{T}-TaS$_2$ provides a prototypical platform, hosting multiple competing charge-density-wave (CDW) phases, a photoinduced hidden state, and superconductivity under doping or pressure~\cite{sipos2008mott,li2012fe,stojchevska2014ultrafast,vaskivskyi2015controlling,vaskivskyi2016fast,stahl2020collapse,wilson1975charge,fazekas1979electrical,yoshida2015memristive}. In its nearly commensurate (NC) and commensurate (C) CDW phases, broken in-plane mirror symmetry gives rise to two degenerate chiral configurations, denoted $\alpha$ and $\beta$~\cite{gao2021chiral,yang2022visualization}. However, the energetic preference for homochiral interlayer stacking stabilizes a predominantly single-domain state in pristine 1\textit{T}-TaS$_2$~\cite{zhao2023spectroscopic}, limiting direct optical control between the two chiralities. Ultrafast excitation instead drives the system through a transient incommensurate (IC) CDW phase, where mirror symmetry is restored and domain walls can nucleate~\cite{haupt2016ultrafast,laulhe2017ultrafast,zong2018ultrafast}.

Engineering a ground state with coexisting chiral domains offers a route to ultrafast control beyond the constrained single-domain landscape of pristine 1\textit{T}-TaS$_2$. Ti substitution provides such a tuning parameter by modifying intra- and interlayer electronic correlations, suppressing the C-CDW phase~\cite{fazekas1979electrical,gao2021chiral}, and stabilizing microscopic coexistence of the $\alpha$-NC and $\beta$-NC domains~\cite{fazekas1979electrical,chen2015influence,gao2021chiral,lacinska2022raman,zhao2023spectroscopic,geng2025filling}. This coexistence may open a nonequilibrium pathway for direct chiral-domain interconversion without first traversing the IC phase. However, the equilibrium distribution of these domains and their ultrafast evolution remain unresolved, requiring measurements that simultaneously resolve doping, momentum and timescale.

Here we combine static and time-resolved X-ray diffraction to investigate chemical and optical control of chiral CDW order in Ti-doped 1\textit{T}-TaS$_2$. We first establish how Ti substitution continuously tunes the equilibrium population and correlation lengths of the two chiral domains. We then show that femtosecond photoexcitation produces an anisotropic and asymmetric redistribution from the dominant toward the minority domain over distinct timescales. The sub-picosecond response is synchronized with a coherent amplitude mode, pointing to a phonon-assisted pathway that differs from the IC-mediated dynamics of pristine 1\textit{T}-TaS$_2$. Together with first-principles calculations, these results support a domain-wall-mediated nonthermal pathway and establish equilibrium free-energy-landscape engineering as a strategy for selecting ultrafast transition pathways in correlated materials.

\begin{figure*}[htbp]
\centering
\includegraphics[width=\textwidth]{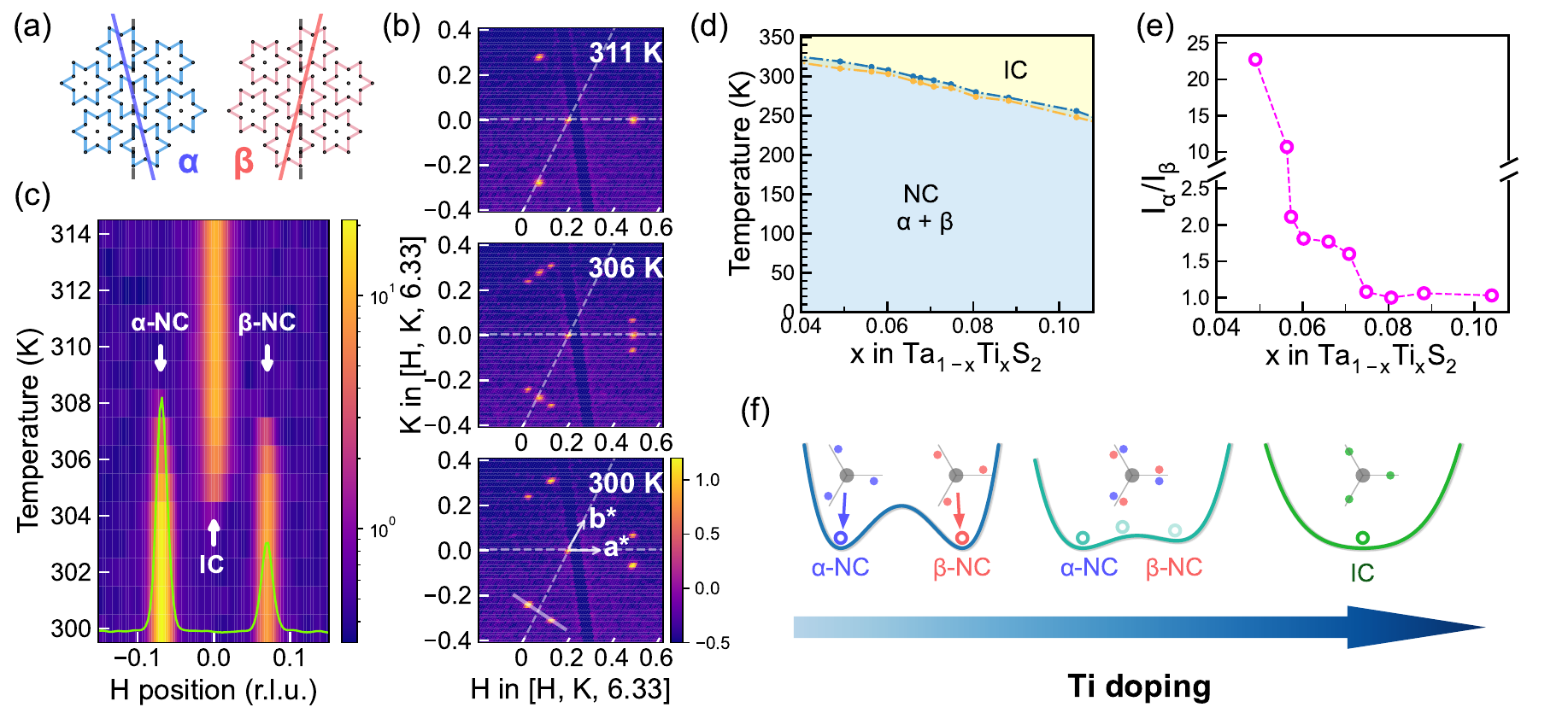}
\caption{\textbf{Ti-doping control of chiral-domain coexistence.}
(a) Schematics of the mirror-related $\alpha$-NC (blue) and $\beta$-NC (red) CDW states. Dots denote Ta atoms, dashed lines mark the crystallographic axes, and solid lines indicate the corresponding NC-CDW wave vectors.
(b) Logarithmic X-ray diffraction intensity maps in the $H$--$K$ plane at $L=6.33$ and selected temperatures. The white line at 300~K marks the cut shown in (c).
(c) Temperature-dependent intensity profiles along this cut during cooling, showing coexisting $\alpha$-NC and $\beta$-NC peaks.
(d) Ti-doping dependence of the NC-CDW onset and IC-CDW disappearance temperatures.
(e) Low-temperature intensity ratio $I_\alpha/I_\beta$ averaged over several tens of diffraction-peak pairs.
(f) Schematic evolution of the free-energy landscape from a predominantly single-chiral state to chiral-domain coexistence and, at higher doping, suppression of the NC-CDW phase.
\label{fig1}}
\end{figure*}

\noindent{\emph{Coexistence of chiral domains}}--To engineer a ground state with coexisting chiral domains, we synthesized 1\textit{T}-Ti$_x$Ta$_{1-x}$S$_2$ single crystals with $x=0.05$--0.11 by chemical vapor transport. Figures~\ref{fig1}(b) and~\ref{fig1}(c) show temperature-dependent X-ray diffraction data for a representative $x=0.06$ sample,  which were performed using a custom-built diffractometer equipped with a Mo K$_{\alpha}$ source with a photon energy of 17.48~keV. The NC-CDW reflections occur at
$Q_\alpha^{\mathrm{NC}}\sim(\sigma_1^{\mathrm{NC}}+\sigma_2^{\mathrm{NC}},-\sigma_2^{\mathrm{NC}},1/3)$ and
$Q_\beta^{\mathrm{NC}}\sim(\sigma_1^{\mathrm{NC}},\sigma_2^{\mathrm{NC}},1/3)$, with $\sigma_1^{\mathrm{NC}}=0.248$ and $\sigma_2^{\mathrm{NC}}=0.068$~\cite{spijkerman1997x}. Upon cooling, the IC--NC transition occurs at $T_{\mathrm{NC}}\sim306$~K, substantially below the $\sim350$~K transition in pristine 1\textit{T}-TaS$_2$~\cite{wilson1975charge}. IC- and NC-CDW reflections coexist over a narrow temperature interval near $T_{\mathrm{NC}}$. In the low-temperature NC-CDW phase, both $\alpha$ and $\beta$ chiral reflections remain visible, in contrast to the predominantly single-domain state of pristine 1\textit{T}-TaS$_2$~\cite{zhao2023spectroscopic}. For convenience, we label the domain with the stronger diffraction intensity as $\alpha$ and the weaker one as $\beta$. Their relative populations are characterized by the integrated-intensity ratio $I_\alpha/I_\beta$. For $x=0.06$, this ratio is approximately $2:1$ (Fig.~\ref{fig1}(c)), demonstrating a stable but imbalanced coexistence of the two chiral domains.

We next tracked $T_{\mathrm{NC}}$ and the chiral-domain imbalance across the doping series. As shown in Fig.~\ref{fig1}(d), $T_{\mathrm{NC}}$ decreases monotonically with increasing Ti concentration, while the IC--NC coexistence interval remains approximately 5--7~K, broader than the 1--3~K range reported for pristine 1\textit{T}-TaS$_2$~\cite{van1980electron,ishiguro1991electron,chen2015influence}. More strikingly, $I_\alpha/I_\beta$ evolves from approximately $20:1$ at $x=0.05$ to nearly $1:1$ for $x\geq0.08$ (Fig.~\ref{fig1}(e)). Ti substitution therefore continuously reduces the imbalance between the two chiral configurations and drives the system toward a macroscopically chirality-balanced state. Within a Landau free-energy picture~\cite{de2021colloquium}, this evolution is consistent with a progressive reduction of the free-energy difference between the $\alpha$ and $\beta$ configurations. Ti doping thus provides a tunable ground state with coexisting chiral domains, creating a favorable starting point for ultrafast optical control.\\

\begin{figure*}[htbp]
\centering
\includegraphics[width=\textwidth]{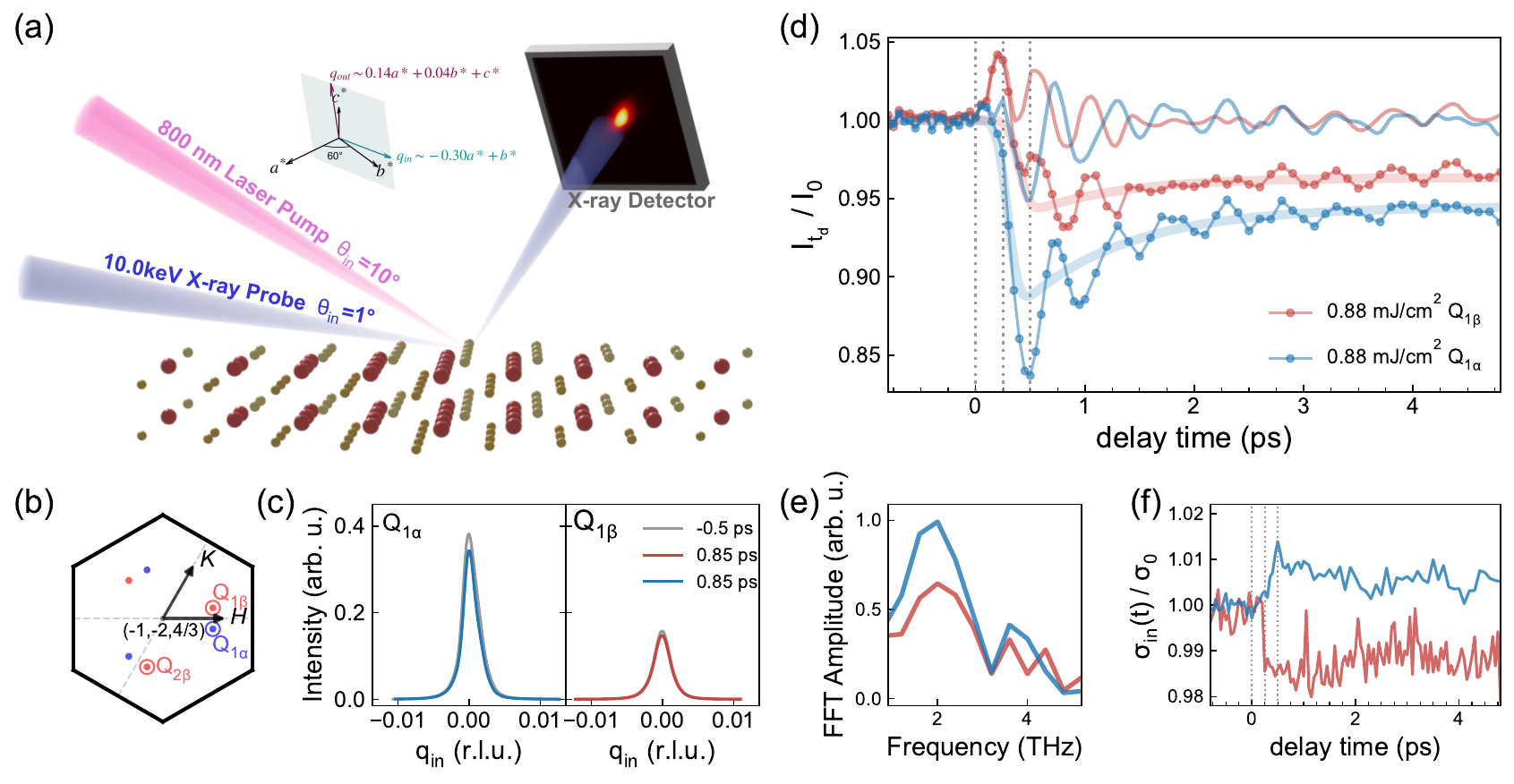}
\caption{\textbf{Ultrafast dynamics of chiral-domain redistribution.}
(a) Schematic of the 800~nm pump--10.0~keV X-ray probe setup. The inset defines the in-plane ($q_{\mathrm{in}}$) and out-of-plane ($q_{\mathrm{out}}$) momentum directions relative to $Q_{1\alpha}$.
(b) Measured CDW reflections near the $(-1,-2,4/3)$ Bragg peak; colored circles mark $Q_{1\alpha}$, $Q_{1\beta}$, and $Q_{2\beta}$.
(c) In-plane profiles of $Q_{1\alpha}$ and $Q_{1\beta}$ before ($t_d=-0.5$~ps) and after ($t_d=0.85$~ps) photoexcitation.
(d) Normalized integrated intensities of $Q_{1\alpha}$ and $Q_{1\beta}$ at an absorbed fluence of 0.88~mJ\,cm$^{-2}$. Thin curves show the oscillatory component and bold curves the nonoscillatory background.
(e) Fourier spectra of (d) showing an amplitude-mode phonon near 2~THz.
(f) Normalized in-plane peak widths $\sigma_{\mathrm{in}}$ extracted from Gaussian fits. Dashed lines mark 0, 250, and 500~fs.
\label{fig2}}
\end{figure*}

\begin{figure*}[htbp]
\centering
\includegraphics[width=0.8\textwidth]{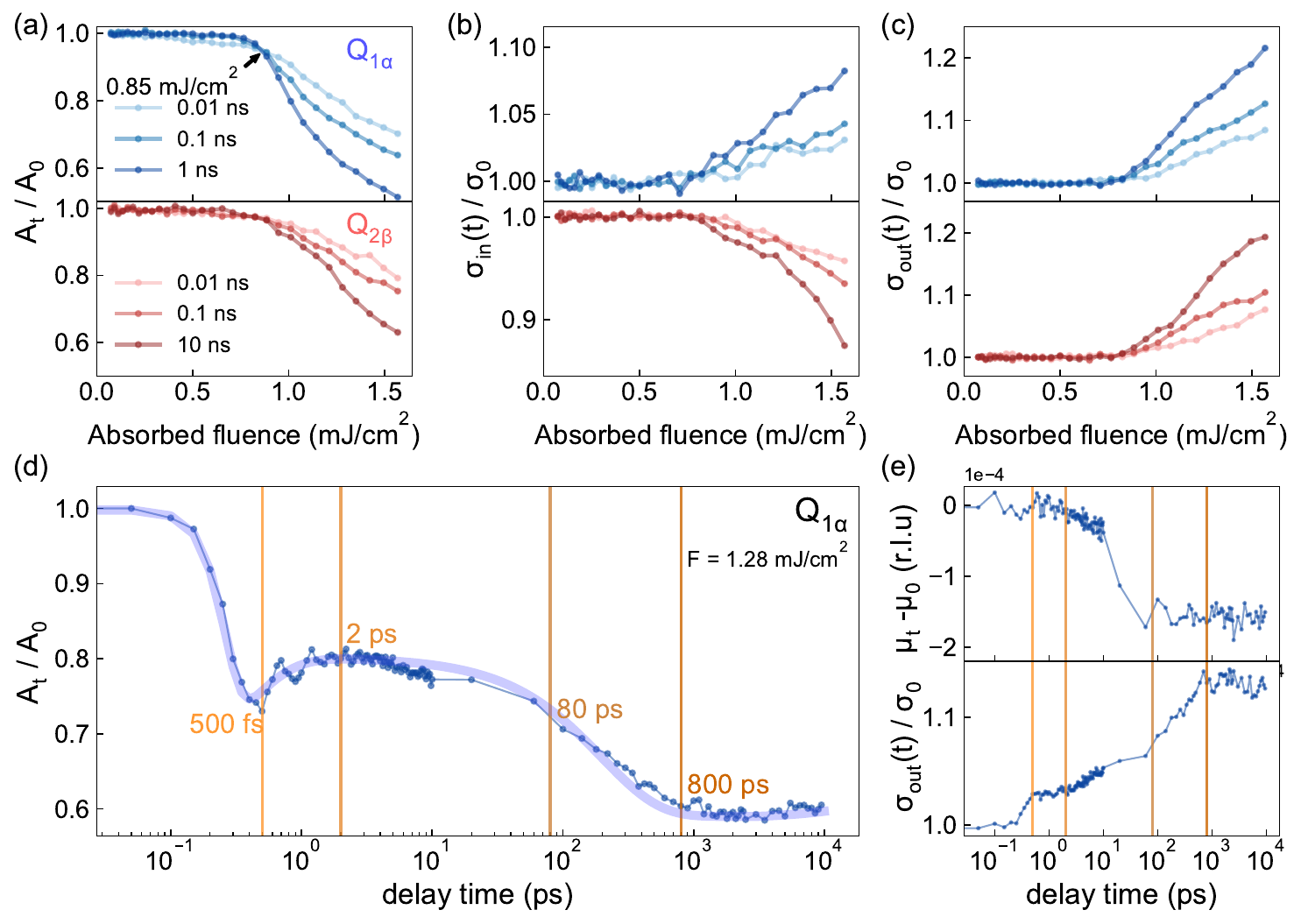}
\caption{\textbf{Anisotropy and timescales of the photoinduced transition.}
(a--c) Fluence dependence of the normalized peak intensity (a), in-plane width (b), and out-of-plane width (c) of the $Q_{1\alpha}$ (top) and $Q_{2\beta}$ (bottom) reflections at selected pump--probe delays.
(d) Normalized $Q_{1\alpha}$ intensity at an absorbed fluence of 1.28~mJ\,cm$^{-2}$, plotted on a logarithmic time axis. Vertical lines mark the characteristic times of 0.5, 2, 80, and 800~ps.
(e) Time evolution of the out-of-plane $Q_{1\alpha}$ peak position (top, in units of $c^*$) and normalized peak width (bottom). Vertical lines indicate the same characteristic times as in (d).
\label{fig3}}
\end{figure*}

\noindent{\emph{Ultrafast chiral-domain redistribution}}--We investigated the nonequilibrium dynamics of coexisting chiral domains in the chemically tuned $x=0.06$ sample using time-resolved X-ray diffraction at the Bernina endstation of SwissFEL~\cite{prat2020compact,ingold2019experimental} (Fig.~\ref{fig2}(a)). The sample was excited by 800~nm laser pulses with a duration of approximately 35~fs. The transient responses of the $\alpha$- and $\beta$-domain NC-CDW reflections, $Q_{1\alpha}$ and $Q_{1\beta}$, were probed using 10~keV X-ray pulses with a duration of approximately 50~fs (Fig.~\ref{fig2}(b)). The pump--probe delay was controlled using a mechanical delay stage and a phase shifter, while shot-to-shot variations in the relative arrival time were corrected using the timing-tool signal. The resulting temporal resolution was approximately 60~fs. A two-dimensional detector simultaneously recorded the transient peak profiles along the in-plane and out-of-plane momentum directions.

Before photoexcitation, the two reflections exhibit different in-plane widths (Fig.~\ref{fig2}(c)). Gaussian fits give full width at half maximum (FWHM, defined as $2\sqrt{2\ln2}\sigma$) values of $0.0027\,a^*$ for the dominant $Q_{1\alpha}$ peak and $0.0029\,a^*$ for the minority $Q_{1\beta}$ peak. Since the inverse peak width tracks the corresponding correlation length, the narrower $Q_{1\alpha}$ reflection indicates that the dominant $\alpha$ domains are, on average, more spatially correlated than the minority $\beta$ domains.

Upon photoexcitation, both CDW intensities are suppressed within $\sim500$~fs and partially recover on the picosecond timescale (Fig.~\ref{fig2}(d)), but with markedly different amplitudes. At an absorbed fluence of 0.88~mJ\,cm$^{-2}$, the $Q_{1\alpha}$ intensity decreases by approximately 10\%, nearly twice the suppression of $Q_{1\beta}$ ($\sim5\%$). Both reflections also exhibit coherent oscillations near 2~THz (Fig.~\ref{fig2}(e)), consistent with the CDW amplitude mode reported previously~\cite{eichberger2010snapshots,zong2018ultrafast,hasaien2025emergent}. The oscillations are initially nearly $\pi$ out of phase, with their relative phase evolving toward zero within $\sim3$~ps.

The peak widths display an equally striking contrast (Fig.~\ref{fig2}(f)). The $Q_{1\beta}$ reflection narrows within $\sim250$~fs, indicating an increase in the $\beta$-domain correlation length, whereas $Q_{1\alpha}$ broadens between $\sim250$ and $\sim500$~fs (Supplementary Fig.~S3). These timescales correspond approximately to half and one full period of the 2~THz coherent phonon, linking the early-time domain reorganization to the lattice dynamics.

The unequal intensity suppression and opposite peak-width evolution demonstrate a domain-selective response rather than uniform melting of the NC-CDW order. In particular, the preferential suppression of the dominant $\alpha$ domains and the simultaneous enhancement of correlations within the minority $\beta$ domains point to a nonthermal reorganization of chiral order~\cite{sun2020transient,sun2020pump}. The initially antiphase coherent oscillations further reveal coupling and competition between the two chiral configurations. Together, these observations show that femtosecond excitation drives a rapid and anisotropic reorganization of the chiral-domain state on a sub-picosecond timescale.\\

\noindent{\emph{Anisotropic domain response and multistage relaxation}}--To determine the fluence threshold for the photoinduced chiral reorganization, we measured the fluence dependence of the $Q_{1\alpha}$ and symmetry-equivalent $Q_{2\beta}$ superlattice reflections at fixed pump--probe delays. As shown in Fig.~\ref{fig3}(a), a threshold appears at $F_{\mathrm{th}}\sim0.85$~mJ\,cm$^{-2}$. Above $F_{\mathrm{th}}$, neither reflection fully recovers within the measured time window of 10~ns. Below threshold, however, the $\beta$-domain intensity recovers within 10~ps, while the $\alpha$-domain reflection remains suppressed. Moreover, the $\alpha$ peak is more strongly suppressed than the $\beta$ peak over the entire fluence range, indicating a persistent domain-selective redistribution during the nanosecond-scale CDW dynamics~\cite{han2015exploration,haupt2016ultrafast,laulhe2017ultrafast}.

The momentum-resolved peak profiles reveal a pronounced anisotropy in this response. Gaussian fits yield the in-plane and out-of-plane widths, $\sigma_{\mathrm{in}}$ and $\sigma_{\mathrm{out}}$, whose inverses track the corresponding correlation lengths. Above $F_{\mathrm{th}}$, the out-of-plane correlation lengths of both domains decrease, with a slightly weaker reduction for the $\beta$ domain (Fig.~\ref{fig3}(c)), consistent with photoinduced screening of interlayer correlations. In contrast, the in-plane correlations evolve oppositely (Fig.~\ref{fig3}(b)): $\xi_{\mathrm{in}}^{(\alpha)}$ decreases, whereas $\xi_{\mathrm{in}}^{(\beta)}$ increases and remains enhanced for up to 10~ns. Thus, photoexcitation weakens interlayer coherence in both domains while fragmenting the dominant $\alpha$ domains and promoting the in-plane growth of $\beta$-domain correlations. This anisotropic evolution is consistent with a long-lived redistribution of chiral order from the dominant toward the minority domain.

We further resolved the relaxation pathway by tracking the $Q_{1\alpha}$ intensity, out-of-plane width, and peak position over 10~ns at an absorbed fluence of 1.28~mJ\,cm$^{-2}$ (Figs.~\ref{fig3}(d,e)). Four dynamical regimes can be distinguished. Within the first 500~fs, electronic excitation and the initial domain-selective reorganization produce rapid changes in the peak intensity and width. Between 500~fs and 2~ps, electron--phonon energy transfer drives a partial recovery of the $\alpha$-domain signal. From 2 to 80~ps, a second suppression develops, consistent with the onset of a photoinduced NC--IC transformation similar to that reported in pristine TaS$_2$~\cite{han2015exploration,haupt2016ultrafast,laulhe2017ultrafast}. Over the same interval, a laser-induced strain wave shifts the diffraction condition, as reflected by the evolution of the peak position $\mu$. Between 80 and 800~ps, out-of-plane thermal diffusion dominates, while the suppression of the NC-CDW signal approaches saturation. Beyond 1~ns, the system gradually cools and recovers, while retaining a modified chiral-domain distribution.

\begin{figure*}[htbp]
\centering
\includegraphics[width=\textwidth]{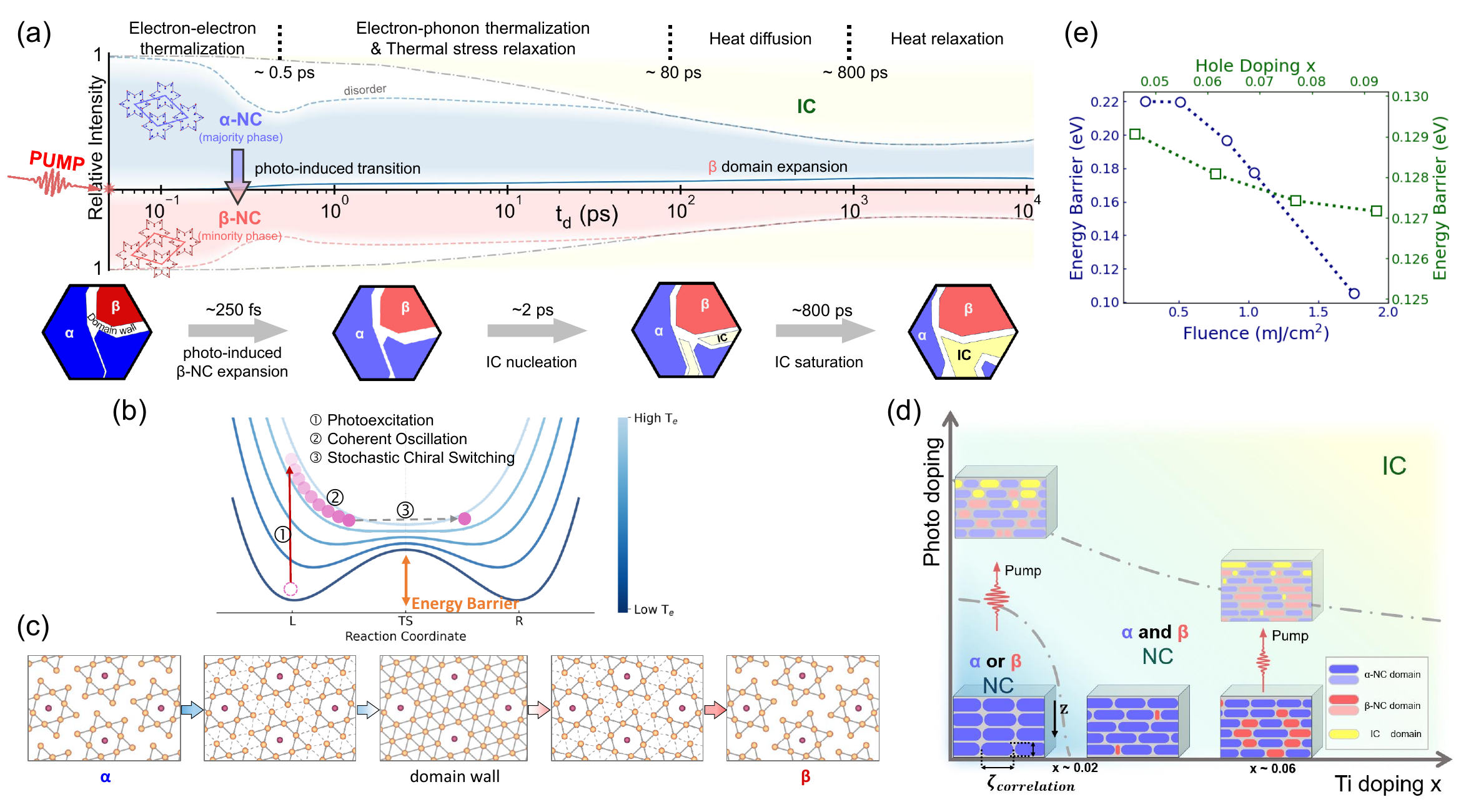}
\caption{\textbf{Mechanism of photoinduced chiral-domain redistribution.}
(a) Schematic timeline of the measured dynamics. The upper panel shows the redistribution of diffraction intensity, and the lower panel illustrates the corresponding in-plane domain evolution.
(b) Schematic free-energy landscape for the light-induced chiral transition.
(c) Proposed real-space pathway involving transient domain-wall configurations.
(d) Comparison of Ti doping and photoexcitation in tuning the chiral-domain population and the in-plane and out-of-plane correlation lengths of the NC- and IC-CDW phases.
(e) Calculated energy difference between the chiral and domain-wall-like configurations as a function of electronic excitation (blue) and hole doping (green). 
\label{fig4}}
\end{figure*}

~\\

\noindent{\emph{Discussion}}--The photoinduced dynamics of Ti-doped 1\textit{T}-TaS$_2$ are summarized in Fig.~\ref{fig4}(a). The in-plane growth of the $\beta$ domain occurs in two regimes: an ultrafast nonthermal response within $\sim250$~fs, followed by a slower NC--IC relaxation extending from $\sim2$~ps to $\sim1$~ns. The latter is consistent with thermally mediated domain redistribution~\cite{sun2020transient,sun2020pump}, whereas the initial response points to a distinct nonthermal pathway. To examine this early-time process, we performed time-dependent density-functional-theory molecular-dynamics simulations~\cite{meng2008real,lian2018photoexcitation,lian2020ultrafast}. The calculations reproduce a coherent CDW amplitude-mode oscillation near 2~THz (Supplementary Fig.~S5), in agreement with experiment. Within the displacive excitation of coherent phonons framework~\cite{zeiger1992theory,mizoguchi2002coherent,wang2025directional}, photoexcitation shifts the potential-energy surface toward a configuration with reduced CDW distortion that is closer to the domain-wall state. During the first half-cycle, the $\alpha$ phase is driven toward this transient configuration, facilitating access to the $\beta$ phase on a timescale set by the coherent phonon (Fig.~\ref{fig4}(b,c)). The calculations therefore support a phonon-assisted channel for chiral-domain reorganization.

The experimental dynamics, however, do not follow a simple sequential conversion. The increase in the $\beta$-domain correlation length is established within $\sim250$~fs, preceding the decrease in the $\alpha$-domain correlation length at $\sim500$~fs (Fig.~\ref{fig2}(f) and Supplementary Fig.~S3). This earlier narrowing of the $\beta$ reflection suggests that the initial reorganization may nucleate within pre-existing domain-wall regions rather than inside the $\alpha$ domains. Such metastable regions have been observed by scanning tunneling microscopy as ring-shaped Star-of-David structures~\cite{geng2025filling}. Although direct confirmation will require ultrafast real-space probes, these regions provide a plausible nucleation pathway for the photoinduced chiral redistribution. The dynamics are also strongly anisotropic. Photoexcitation suppresses out-of-plane coherence in both domains, consistent with screening of the weaker interlayer correlations, while the in-plane response is domain selective: the dominant $\alpha$ domains fragment as the minority $\beta$ domains expand. This contrast reflects the quasi-two-dimensional character of the CDW correlations~\cite{su2012collective} and shows that the chiral reorganization occurs primarily within the TaS$_2$ layers.

The equilibrium and nonequilibrium results further reveal a close correspondence between chemical and optical tuning (Fig.~\ref{fig4}(d)). Ti substitution progressively balances the populations of the two chiral domains, whereas photoexcitation preferentially suppresses the initially dominant $\alpha$ domains and promotes the relative growth of the $\beta$ domains. Both perturbations therefore rebalance the competition between chiral orders, although they act through distinct microscopic mechanisms and timescales. Their common anisotropic evolution--reduced out-of-plane coherence and opposing in-plane trends--suggests that static chemical tuning and ultrafast optical excitation act on the same underlying competition between intra- and interlayer correlations.

The calculations provide further support for this picture (Fig.~\ref{fig4}(e)). Increasing the electronic temperature reduces the energy difference between the chiral and domain-wall-like configurations, indicating that photoexcitation facilitates access to the transient domain-wall pathway~\cite{huang2024ultrafast}. Hole doping also reduces this energy difference~\cite{geng2025filling}, but less strongly, suggesting that carrier doping alone is insufficient to account for the observed redistribution and that local effects associated with Ti substitution, such as orbital reconstruction, are also important~\cite{gao2021chiral,van2011chirality}.

Taken together, our results establish a cooperative route for controlling chiral order in a correlated material. Chemical substitution prepares a tunable landscape of coexisting chiral domains, while femtosecond excitation transiently reshapes this landscape and drives an anisotropic, phonon-assisted redistribution from the dominant toward the minority chirality. More broadly, equilibrium landscape engineering provides a strategy for selecting nonequilibrium pathways in systems with competing or intertwined order parameters, including multiferroics~\cite{eerenstein2006multiferroic,kimel2005ultrafast,bustamante2025ultrafast}.

~\\
\noindent{\emph{Acknowledgments}}--Y. P. is grateful for financial support
from the Beijing Natural Science Foundation
(Grant No. JQ24001), the Ministry of Science and
Technology of China (Grants No. 2024YFA1408702 and
No. 2021YFA1401903), and the National Natural Science
Foundation of China (Grant No. 12374143). S. M. is grateful for financial support from the National Natural Science Foundation of China (Grant No.
12450401), the
Ministry of Science and Technology of China (Grant No.
2021YFA1400200). We acknowledge the Paul Scherrer Institut, Villigen, Switzerland, for provision of beamtime at the Bernina beamline of SwissFEL.
}

~\\

~\\

\end{document}